\documentclass[12pt, draftclsnofoot, onecolumn]{IEEEtran}
\usepackage{amsmath}
\usepackage{amssymb}
\usepackage{graphicx}
\usepackage{verbatim}
\usepackage{cite}

\newcommand{\sir}{{\rm SIR}}

\newcommand{\ex}{{\rm e}}
\newcommand{\dd}{{\rm d}}
\newcommand{\ie}{{\em i.e.}}
\newcommand{\eg}{{\em e.g.}}
\newcommand{\ea}{{\em et al. }}
\newcommand{\adhoc}{{\em ad hoc} }

\newcommand{\E}{\mathbb{E}}

\begin{document}
\title{An overview of the transmission capacity of wireless networks}

\author{Steven Weber, Jeffrey G. Andrews, Nihar Jindal
\thanks{S.~Weber is with Drexel University, J.G.~Andrews is with the University of Texas at Austin, N.~Jindal is with the University of Minnesota.  The contact author is N.~Jindal ({\sf nihar@ece.umn.edu}).  This research was supported by NSF grant no. 0635003 (Weber), nos. 0634979 and 0643508 (Andrews), no. 0634763 (Jindal), and the DARPA IT-MANET program, grant no. W911NF-07-1-0028 (all authors).  Manuscript date: \today.}}

\maketitle

\begin{abstract}
This paper surveys and unifies a number of recent contributions that have collectively developed a metric for decentralized wireless network analysis known as transmission capacity.  Although it is notoriously difficult to derive general end-to-end capacity results for multi-terminal or \adhoc networks, the transmission capacity (TC) framework allows for quantification of achievable single-hop rates by focusing on a simplified physical/MAC-layer model.  By using stochastic geometry to quantify the multi-user interference in the network, the relationship between the optimal spatial density and success probability of transmissions in the network can be determined, and expressed -- often fairly simply -- in terms of the key network parameters. The basic model and analytical tools are first discussed and applied to a simple network with path loss only and we present  tight upper and lower bounds on transmission capacity (via lower and upper bounds on outage probability).  We then introduce random channels (fading/shadowing) and give TC and outage approximations for an arbitrary channel distribution, as well as exact results for the special cases of Rayleigh and Nakagami fading.  We then apply these results to show how TC can be used to better understand scheduling, power control, and the deployment of multiple antennas in a decentralized network.  The paper closes by discussing shortcomings in the model as well as future research directions.
\end{abstract}

\section{\label{sec:intro}Introduction}

This paper presents the recently developed framework for the outage probability and \emph{transmission capacity} \cite{WebYan2005} in a one hop wireless \adhoc network.  The transmission capacity is defined as the number of successful transmissions taking place in the network per unit area, subject to a constraint on outage probability. In addition to being of general interest, the advantange of transmission capacity -- relative to, say, the transport capacity or average sum throughput -- lies largely in that it can be exactly derived in some important cases, and tightly bounded in many others, as we shall show.  From the expressions and approach given in this paper the exact dependence between system performance (transmission capacity, outage probability) and the possible design choices and network parameters are laid bare. In contrast to the proposed framework, nearly all other work on \adhoc networks must resort to scaling laws or numerical simulations, in which case intuition and/or precision is usually lost.

The first goal of this paper is to concisely summarize the new analytical tools (largely drawn from the field of stochastic geometry \cite{StoKen1996,Kin1993}) that have been developed over numerous papers by the authors and others.  Because these techniques have been developed somewhat independently depending on the problem of interest, the system model in \S\ref{sec:sysmodel} applied to the baseline model of pathloss attenuation without fading in \S\ref{sec:model} will help newcomers to the area understand the various approaches in context.

The second goal is to show how this framework can be used to give crisp insights into wireless network design problems.  In the past few years, the transmission capacity approach has been applied to various design problems by a growing group of researchers (see \cite{WebYan2005,StaPro2007b, BloJin2009,HuaLau2009,YinGao2009, LouMcK2009}).  Although transmission capacity was originally developed to analyze spread spectrum in ad hoc networks, it has proven to be a metric with considerable breadth of application.  Since decentralized wireless networks are generally very difficult to characterize, the intuitive and simple-to-compute qualities of transmission capacity have made it a popular choice for a large number of possible systems, including: $i)$ direct-sequence and frequency-hopping spread spectrum \cite{WebYan2005,AndWeb2007,StaPro2007b}, $ii)$ interference cancellation \cite{WebAnd2007a,BloJin2009}, $iii)$ spectrum sharing in unlicensed, overlaid, and cognitive radio networks \cite{JinAnd2008,HuaLau2009,YinGao2009,ChaAnd2009b}, $iv)$ scheduling \cite{WebAnd2007a} and power control \cite{WebAnd2007b,JinWeb2008}, $v)$ and the use of multiple antennas (which had resisted characterization by other methods) \cite{HunAnd2008a,LouMcK2009, HuaAndSub,StaPro2007a,KouAnd2009a,JinAnd2009,HunAnd2008b,VazHea10}.  Other researchers have also further studied the basic tradeoffs between outage probability, data rate, and transmission capacity for general networks \cite{Hae2009}.
We selectively discuss some of these applications.  \S\ref{sec:Design2} addresses networks with fading channels, with a focus on Rayleigh (\S\ref{ssec:Rayleigh}) and Nakagami (\S\ref{ssec:Nakagami}) fading, scheduling (\S\ref{ssec:scheduling}), and power control (\S\ref{ssec:FPC}).  \S\ref{sec:MIMO} addresses the use of multiple antennas, with discussions of diversity (\S\ref{ssec:diversity}), spatial interference cancellation (\S\ref{ssec:cancel}), and spatial multiplexing (\S\ref{ssec:mux}).

The third goal of the paper is to stimulate new efforts to further the tools presented here, both in making them more general and in applying them to new problems. We readily concede that the presented model has some nontrivial shortcomings at present, and we identify those as well as possible avenues forward in \S\ref{sec:future}.

\section{\label{sec:sysmodel}System model}

We introduce the system model in \S\ref{ssec:2a}, discuss relevant mathematical background in \S\ref{ssec:2b}, and elaborate on the connection with transport capacity in \S\ref{ssec:2c}.

\subsection{\label{ssec:2a}Mathematical model and assumptions}

We consider an \adhoc wireless network consisting of a large (infinite) number of nodes spread over a large (infinite) area.  The network is {\em uncoordinated}, meaning transmitters do not coordinate with each other in making transmission decisions.  That is, nodes employ Aloha \cite{Abr1970} (\ie, in each slot, each node independently decides whether to transmit or to listen) as the medium access control (MAC) protocol.  We view the network at a snapshot in time, where the locations of the transmitting nodes at that snapshot are assumed to form a stationary Poisson point process (PPP) on the plane of intensity $\lambda$, denoted $\Pi(\lambda) = \{X_i\}$, where each $X_i \in \mathbb{R}^2$ is the location of  interfering transmitter $i$.  The PPP assumption for node locations is valid when the uncoordinated transmitting nodes are independently and uniformly distributed over the network arena, which is often reasonable for networks with indiscriminate node placement or substantial mobility.  If intelligent transmission scheduling is performed, the resulting transmitter locations will most certainly not form a PPP, so this paper's analytical framework is primarily applicable to uncoordinated transmitters.  Although suboptimal, such a model may be reasonable in cases where the overhead associated with scheduling is prohibitively high, for example due to highly mobile nodes, bursty traffic, or rigid delay constraints.  We also note that this framework has been extended to CSMA, and the gains are not that large over Aloha \cite{BacBla2006,HasAnd2007}.  Viewing the network at a single snapshot in time restricts our focus to characterizing the performance of one-hop transmissions with specified destinations.  That is, our attention is on (uncoordinated) MAC layer performance, but our model neither addresses nor precludes any multi-hop routing scheme.  These model limitations are further discussed in \S\ref{sec:future}.

\begin{figure}
\centering
\includegraphics[width=2.5in]{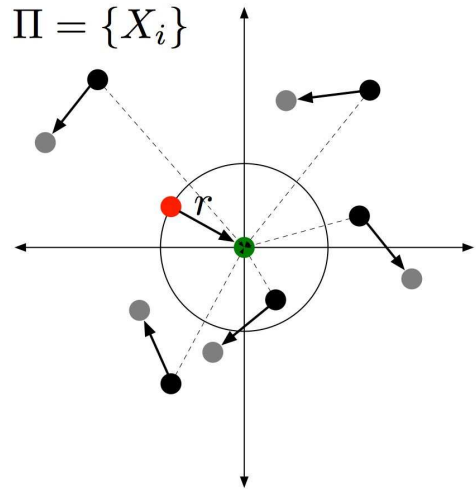}
\caption{The transmitter locations (black circles) at a typical time form a Poisson process, $\Pi$; each transmitter has an assigned receiver (gray circles) located at distance $r$.  The reference communications link has a reference receiver at the origin (green) and a reference transmitter at distance $r$ (red).  Each black transmitter generates interference seen at the reference receiver, indicated by the dashed lines.}
\label{fig:a}
\end{figure}

Each transmitter is assumed to have an assigned receiver at a fixed distance $r$ (meters) away.  This assumption may be easily relaxed (\eg, see \cite{WebAnd2007b} and \cite{JinWeb2008}) but at the cost of complicating the derived expressions without providing additional insight.  The set of receivers is disjoint with the set of transmitters.  Because the network is infinitely large and spatially homogeneous, the statistics of $\Pi(\lambda)$ are unaffected by the addition of a placed transmitter and receiver pair, and, more importantly, this pair is ``typical'' in that the performance experienced at the reference pair characterizes the node-average performance in the network  (Slivnyak's Theorem \cite{StoKen1996}).  Without loss in generality we place the reference receiver at the origin ($o$), and the reference transmitter is located $r$ meters away. See Fig.~\ref{fig:a}.  Note that the locations of the other receivers are not important because the reference receiver's performance only depends upon the positions of the transmitters.

Each transmitter is usually assumed to employ unit transmission power (except when we discuss power control in \S\ref{ssec:FPC}).  The channel strength is assumed to be solely determined by pathloss and fading, \ie, the received power at distance $d$ is $H d^{-\alpha}$, where $\alpha > 2$ is the pathloss exponent and $H$ is the fading coefficient.  All fading coefficients are assumed to be independent and identically distributed (iid).  This simplified model has been shown to capture the key distance dependency in \adhoc networks, and minor alterations to it such as adding an attenuation constant or forcing the received power to be less than one increase the analytical complexity with little apparent benefit \cite{InaChi2009}.  We study networks without fading ($H=1$) in \S\ref{sec:model} then with fading in \S\ref{sec:Design2}.

We treat interference as noise, assume that the ambient/thermal noise is negligible, and assert transmission success to be determined by the signal to interference plus noise ratio (SINR) lying above a specified threshold $\beta$.  The assumption of negligible thermal noise may be easily relaxed (\eg, see \cite{WebAnd2007b} and \cite{JinWeb2008}) but at the cost of complicating the derived expressions without providing additional insight.  The {\em outage probability} (OP), denoted by $q$, is the probability that the signal to interference ratio (SIR) at the reference receiver is below a specified threshold $\beta$ required for successful reception:
\begin{equation}
\label{eqn:1}
q(\lambda) \equiv \mathbb{P}(\sir < \beta) = \mathbb{P} \left( \frac{S r^{-\alpha}}{\sum_{i \in \Pi(\lambda)} I_i |X_i|^{-\alpha}} < \beta \right) = \mathbb{P} \left( Y > \frac{1}{\beta} \right),
\end{equation}
where $Y \equiv \frac{1}{S r^{-\alpha}} \sum_{i \in \Pi(\lambda)} I_i |X_i|^{-\alpha}$ is defined as the aggregate interference power seen at the reference receiver at the origin, normalized by the signal power $S r^{-\alpha}$.  The last expression in (\ref{eqn:1}) highlights the fact that, conditioned on $S$, the OP is the tail probability of the aggregate interference level expressed as a shot noise process.

The randomness is in the interferer locations, $\{X_i\}$, and the fading coefficients, $S$ and $\{I_i\}$.  The OP is a function of $\alpha,\beta,\lambda,r$ and the fading statistics.  Note that $q$ is continuous monotone increasing in $\lambda$ and is onto $[0,1]$. Our primary performance metric is the {\em transmission capacity} (TC) which takes a target OP $\epsilon$ as a parameter:
\begin{equation}
\label{eqn:2}
c(\epsilon) \equiv q^{-1}(\epsilon)(1-\epsilon), ~~ \epsilon \in (0,1).
\end{equation}
It is the spatial intensity of attempted transmissions $q^{-1}(\epsilon)$ associated with OP $\epsilon$, thinned by the probability of success, $1-\epsilon$.  The quantity $\epsilon$ is a network-wide quality of service measure, ensuring a typical attempted transmission will succeed with probability $1-\epsilon$.  The transmission capacity has units of number of transmission attempts per unit area, \ie, it is a measure of spatial intensity of transmissions.  Note that the OP $q(\lambda)$ is defined for an arbitrary transmission intensity $\lambda$, and $c(\epsilon)$ is simply that value of $\lambda$ such that $q(c(\epsilon)/(1-\epsilon)) = \epsilon$.  The definition of TC is motivated by several factors: $i)$ fixing the OP at $q =\epsilon$ is a useful and simple, albeit coarse, characterization of network performance, $ii)$ the TC is tractable and can be computed, or at least bounded, for many useful network design questions.  A summary of the mathematical notation employed in this paper is given in Table~\ref{tab:1}.

\subsection{\label{ssec:2b}Mathematical background}

The key underlying mathematical concept is the shot-noise process first developed in 1918 \cite{Sch1918},
\begin{equation}
\label{eqn:d} Y(t) = \sum_{j = -\infty}^{\infty} h(t-t_j),
\end{equation}
where $\{t_j\}$ is a stationary Poisson point process (PPP) on $\mathbb{R}$ and $h(t)$ is a (linear, time-invariant) impulse response function \cite{Ric1944,Gub1996}.  Here $Y(t)$ is the superposition of all signals, appropriately attenuated to time $t$. If we instead interpret $\{t_j\}$ as locations on the plane, $t$ as the location of a reference receiver, $h(t)$ as a channel attenuation function, and $t-t_j$ as the distance from $t_j$ to $t$, then $Y(t)$ may be interpreted as the cumulative interference power seen at $t$.  A power-law impulse response, $h(t) = Kt^{-\alpha}$ \cite{LowTei1990} makes the process $\{Y(t)\}$ L\'{e}vy stable \cite{ShaNik1993}.

The use of spatial models in wireless communications dates back to the late 1970's \cite{KleSil1978,MusWas1978}.  There was in fact quite extensive work on the model in which nodes are located according to a 2-D PPP, Aloha is used, a routing protocol determines the node for which each transmitted packet is intended for, and the received SINR and specifics of the communication protocol determine conditions for transmission success; see \cite{KleSil1987} for an overview of early results.  The aggregate interference process in an \adhoc network was first recognized as L\'{e}vy stable in \cite{Sou1990,SouSil1990,Sou1992}, and its characteristic function was studied in \cite{IloHat1998}.  A series of papers by Baccelli {\em et al.}
demonstrated the power of stochastic geometry for modeling a wide range of problems within wireless communications, as
summarized in \cite{BacBlaNOW,FraMee2007}.

We note that there have been several very helpful tutorials on applying stochastic geometry to wireless networks developed in the last year,
including the comprehensive two-volume monograph by Baccelli \cite{BacBlaNOW}, a monograph by Ganti and Haenggi that has many of the available results on non-homogeneous Poisson node distributions \cite{HaeGanNOW}, a summary tutorial article for a JSAC special issue on the topic \cite{HaeAnd2009}, and a tutorial by Win \ea on characterizing interference in Poisson fields \cite{WinPin2009}.  We refer readers to those references (and \cite{Kin1993,StoKen1996}) for background.

\subsection{\label{ssec:2c}Relationship to transport capacity}

The general subject of the paper is the analysis of capacity and outage probability of wireless \adhoc networks.  Ideally, one could determine the capacity region of an \adhoc network, which would be the set of maximum rates that could be achieved simultaneously between all possible pairs in the network, and hence is $n(n-1)$ dimensional for $n$ (full-duplex) users.  Even if this was obtainable -- which it has not been despite considerable efforts \cite{TouGol2003} -- it would still likely not capture some key aspects of an \adhoc network, which call for information to be moved over space. Gupta and Kumar pioneered an important line of work on \emph{transport capacity} in \cite{GupKum2000}, which measures the end-to-end sum throughput of the network multiplied by the end-to-end distance. Representative publications include \cite{XueKum2006,XieKum2004,JovVis2004,LevTel2005,FraDou2007}.  A key feature of all these works is that it is not possible to compute the exact transport capacity in terms of the system parameters, and although bounds and closed-form expressions are available in some cases, the best-known results are stated in the form of \emph{scaling laws} that quantify how the volume of the capacity region grows with the number of nodes in the network. The most accepted conclusion is that the capacity grows sublinearly as $\Theta(\sqrt{n})$, which can be achieved with multi-hop transmission and treating multi-user interference as noise, as proven in several different ways \cite{GupKum2000,LevTel2005,TouGol2004} including recently using Maxwell's equations \cite{FraMig2009}.  Generous assumptions on mobility \cite{GroTse2002}, bandwidth \cite{NegRaj2004}, or cooperation \cite{OzgLev2007} result in more optimistic scaling laws.

The transport capacity, $C_T(n)$, is defined as the maximum distance-weighted sum rate of communication over all pairs of $n$ nodes \cite{GupKum2000}.  In an extensive network, where the density of nodes per unit area is constant, the transport capacity has been shown to grow as $C_T(n) = \Theta(n)$ as $n \to \infty$, with units of bit-meters per second \cite{XueKum2006}.  Roughly speaking, there can be $\Theta(n)$ simultaneous nearest-neighbor transmissions in the network, and the distance and the rate of communication between nearest-neighbors are both $\Theta(1)$, yielding $C_T(n) = \Theta(n)$.

Comparison of transport capacity and TC is facilitated by normalizing the transport capacity by the network area, $A(n) = \Theta(n)$, giving $C_T(n)/A(n) = \Theta(1)$ in units of bit-meters per second per unit area.  Within the TC framework, assuming communication at the Shannon rate of $\log_2(1 + \beta)$, this metric also is $\Theta(1)$ and is precisely $c(\epsilon)\log_2(1+\beta)r$. Thus, transmission and transport capacity are consistent in the scaling sense.  Furthermore, by abstracting out the end-to-end and multihop aspect of the network, the transmission capacity framework allows for a detailed study of the critical constant term; this is generally very difficult to do if using transport capacity.   Transport capacity and TC are complementary metrics: transport capacity gives order optimal throughput, optimized over all MAC and routing techniques, while TC gives detailed performance and design insights for the lower layers of the network.

\section{\label{sec:model} Baseline model: Path loss only}
In this section, a baseline model is presented where the only randomness is in the position of the nodes, \ie, there is no fading ($S = 1$ and $I_i = 1$ for each $i$ in (\ref{eqn:1})).  Upper and lower bounds are given on outage probability and transmission capacity, emphasizing the impact that \emph{dominant} (strong) interferers have on the sum of the interference. The impact of fading is addressed in \S\ref{sec:Design2}.

\subsection{\label{ssec:3a} Exact results}

The points of the 2-D PPP of intensity $\lambda$, i.e., $\Pi(\lambda) = \{X_i\} \subset \mathbb{R}^2$, may be mapped to a 1-D PPP of unit intensity using Corollary 2 in \cite{Hae2005}.  In particular, $\pi \lambda |X_i|^2 \sim T_i$, where $|X_i|^2$ is the squared distance from the origin of the $i^{th}$ nearest transmitter, and $T_i$ is the distance from the origin of the $i^{th}$ nearest point in a unit intensity 1-D PPP.  Applying this to the normalized interference power $Y$ in (\ref{eqn:1}) gives:
\begin{equation}
Y = r^{\alpha} \sum_{i \in \Pi(\lambda)} |X_i|^{-\alpha} = (\pi \lambda)^{\frac{\alpha}{2}} r^{\alpha} \sum_{i \in \Pi(\lambda)} (\pi \lambda |X_i|^2)^{-\frac{\alpha}{2}} = (\pi r^2 \lambda)^{\frac{\alpha}{2}} \sum_{i \in \Pi_1(1)} T_i^{-\frac{\alpha}{2}},
\end{equation}
where the notation $\Pi_1(1)$ indicates a 1-D PPP of intensity $1$.  The corresponding OP in (\ref{eqn:1}) becomes
\begin{equation}
q(\lambda) = \mathbb{P} \left( (\pi r^2 \lambda)^{\frac{\alpha}{2}} \sum_{i \in \Pi_1(1)} T_i^{-\frac{\alpha}{2}} > \frac{1}{\beta} \right) = \mathbb{P} \left( Z_{\alpha} > \frac{1}{ (\pi r^2 \lambda)^{\frac{\alpha}{2}} \beta} \right) = \bar{F}_{Z_{\alpha}} \left( \left( (\pi r^2 \lambda)^{\frac{\alpha}{2}} \beta \right)^{-1} \right),
\end{equation}
where $Z_{\alpha} \equiv \sum_{i \in \Pi_1(1)} T_i^{-\frac{\alpha}{2}}$ is a random variable whose distribution depends only on $\alpha$
 and $\bar{F}_{Z_{\alpha}} (\cdot)$ is the complementary cumulative distribution function (CCDF) of  $Z_{\alpha}$.
  Using $\bar{F}_{Z_{\alpha}}^{-1}(\cdot)$ to denote the inverse, and solving $\bar{F}_{Z_{\alpha}} \left( \left( (\pi r^2 \lambda)^{\frac{\alpha}{2}} \beta \right)^{-1} \right) = \epsilon$ for $\lambda$ allows the TC to be written as:
\begin{equation}
\label{eq-transcap_implicit}
c(\epsilon) = \frac{ \left(\bar{F}_{Z_{\alpha}}^{-1}(\epsilon)\right)^{-\frac{2}{\alpha}} (1 - \epsilon) }{\pi r^2 \beta^{\frac{2}{\alpha}}}.
\end{equation}
These transformations highlight that the essential difficulty in computing the OP and the TC lies in computing the distribution of the stable rv $Z_{\alpha}$.

In fact the only $\alpha > 2$ for which $Z_{\alpha}$ has a distribution expressible in closed-form is for $\alpha=4$, which is the inverse Gaussian distribution.  Important early results for this special case are due to Sousa and Silvester \cite{SouSil1990} (Eqn. (21)).  In particular, they give an exact expression for the OP in terms of the CDF of the standard normal rv, $Q(z) = \mathbb{P}(Z \leq z)$, for $Z \sim N(0,1)$:
\begin{equation}
\label{eqn:11}
q(\lambda) = 2 Q \left(\sqrt{\pi/2} \lambda \pi r^2 \sqrt{\beta}  \right) - 1.
\end{equation}
The corresponding {\em exact} expression for the TC is:
\begin{equation}
\label{eqn:12}
c(\epsilon) = \frac{\sqrt{2/\pi}(1-\epsilon) Q^{-1}\left((1+\epsilon)/2 \right)}{\pi r^2 \sqrt{\beta}} .
\end{equation}
An additional exact result is given for the case of Rayleigh fading in \S\ref{ssec:fading}.  The general unavailability of closed form expressions for the distribution of $Z_{\alpha}$ motivates the search for lower and upper bounds, which we discuss next.

\subsection{\label{ssec:LB} Lower outage bound: dominant nodes}

A lower bound on the probability of outage is obtained by partitioning the set of interferers $\Pi$ into dominating and non-dominating nodes.  A node $i$ is dominating if its interference contribution alone is sufficient to cause outage at the receiver.  We call dominating nodes near (n) nodes and non-dominating nodes far (f) because because dominating nodes must be within some distance of the origin, and non-dominating nodes must be far from the origin.  The dominating nodes may be defined geometrically as the interferers located inside a disk centered at the origin of radius $\beta^{\frac{1}{\alpha}}r$:
\begin{equation}
\label{eqn:3}
\Pi^{\rm n}(\lambda) \equiv \left\{X_i : \frac{r^{-\alpha}}{|X_i|^{-\alpha}} < \beta \right\} ~ = ~ \left\{X_i : |X_i| < \beta^{\frac{1}{\alpha}}r \right\} = \Pi(\lambda) \cap b \left(o,\beta^{\frac{1}{\alpha}}r \right).
\end{equation}
Here $b(o,d) = \{ x \in \mathbb{R}^2 : \|x\| \leq d\}$ denotes the ball centered at the origin $o$ of radius $d$.  The aggregate interference, normalized by the received signal power $r^{-\alpha}$, may be split into aggregate dominant and aggregate non-dominant interference:
\begin{equation}
\label{eqn:4}
Y \equiv \frac{1}{r^{-\alpha}} \sum_{i \in \Pi(\lambda)} |X_i|^{-\alpha}, ~~ Y^{\rm n} \equiv \frac{1}{r^{-\alpha}} \sum_{i \in \Pi^{\rm n}(\lambda)} |X_i|^{-\alpha}, ~~ Y^{\rm f} \equiv \frac{1}{r^{-\alpha}} \sum_{i \not\in \Pi^{\rm n}(\lambda)} |X_i|^{-\alpha},
\end{equation}
where $Y = Y^{\rm n} + Y^{\rm f}$.  The lower bound is obtained by ignoring the non-dominant interference:
\begin{equation}
\label{eqn:5}
q(\lambda) = \mathbb{P}\left( Y^{\rm n} + Y^{\rm f} > \frac{1}{\beta} \right) > \mathbb{P}\left( Y^{\rm n} > \frac{1}{\beta} \right) \equiv q^l(\lambda).
\end{equation}
Note that, by construction, the event $\{Y^{\rm n} > \frac{1}{\beta}\}$ is the same as the event $\{\Pi^{\rm n}(\lambda) \neq \emptyset\}$, which is simply the complement of a void probability for a Poisson process:
\begin{equation}
\label{eqn:6}
q^l(\lambda) = 1 - \mathbb{P}(\Pi^{\rm n}(\lambda) = \emptyset) = 1 - \ex^{- \lambda \left| b \left(o,\beta^{\frac{1}{\alpha}}r \right) \right|} = 1 - \ex^{-\lambda \pi r^2 \beta^{\frac{2}{\alpha}}}.
\end{equation}
By solving $q^l(\lambda) = \epsilon$ for $\lambda$ we get an upper bound on $q^{-1}(\epsilon)$, which yields a TC upper bound:
\begin{equation}
\label{eqn:7}
c^u(\epsilon) = \frac{(1-\epsilon) \log(1-\epsilon)^{-1}}{\pi r^2 \beta^{\frac{2}{\alpha}}} = \frac{1}{\pi \left( \frac{ r \beta^{\frac{1}{\alpha}} }{\sqrt{\epsilon} } \right)^2} + O(\epsilon^2) \mbox{ as } \epsilon \to 0.
\end{equation}
The right hand side is obtained by observing that the first order Taylor series expansion of $(1-\epsilon)\log(1-\epsilon)^{-1}$ around $\epsilon = 0$ equals $\epsilon + O(\epsilon^2)$, where $O(\cdot)$ is the standard ``big-oh'' notation \cite{GraKnu1994}.  Neglecting the $O(\epsilon^2)$ term gives an error $\epsilon - (1-\epsilon)\log(1-\epsilon)^{-1} \approx 0.005$ for $\epsilon = 0.1$.  The right hand side may be interpreted as a disk packing statement.  In particular, the maximum number of transmissions per square meter for fixed $\alpha,\beta,\epsilon,r$ is found by packing disks of radius $R(\alpha,\beta,\epsilon,r) \equiv \frac{r \beta^{\frac{1}{\alpha}}}{\sqrt{\epsilon}}$, each disk with a single transmitter at the center.  This radius clarifies the dependence of the supportable density of transmissions on these four key model parameters.

\subsection{\label{ssec:UB1} Upper outage bounds: Markov, Chebychev, and Chernoff bounds}

We may decompose the outage event in \eqref{eqn:5} as:
\begin{equation}
\label{eq:UB0}
q(\lambda) = \mathbb{P} \left( \left\{ Y^{\rm n} > \frac{1}{\beta} \right\} \cup \left\{ Y^{\rm f} > \frac{1}{\beta} \right\} \cup \left\{ Y^{\rm n} \leq \frac{1}{\beta}, Y^{\rm f} \leq \frac{1}{\beta}, Y^{\rm n}+Y^{\rm f}>\frac{1}{\beta} \right\} \right).
\end{equation}
In words: the event $\{Y^{\rm n}+Y^{\rm f} > 1/\beta\}$ means either $Y^{\rm n}$ or $Y^{\rm f}$ individually exceed $1/\beta$, or they are both below $1/\beta$ but their sum exceeds $1/\beta$.  By construction, however, the event $\{Y^{\rm n} \leq 1/\beta\}$ is the same as the event $\{Y^{\rm n} = 0\}$, which means the third event in (\ref{eq:UB0}) is null.  The probability of the remaining first two events may be written as:
\begin{equation}
\label{eq:UB1}
q(\lambda) = \mathbb{P}\left( Y^{\rm n} > \frac{1}{\beta} \right) + \mathbb{P}\left( Y^{\rm f} > \frac{1}{\beta} \right) - \mathbb{P}\left( Y^{\rm n} > \frac{1}{\beta} \right) \mathbb{P}\left( Y^{\rm f} > \frac{1}{\beta} \right) = q^l(\lambda)+(1-q^l(\lambda)) \mathbb{P}\left( Y^{\rm f} > \frac{1}{\beta} \right),
\end{equation}
where we have exploited the independence of $Y^{\rm n},Y^{\rm f}$ and applied the definition of $q^l(\lambda)$ in (\ref{eqn:5}).  Substituting (\ref{eqn:6}) for $q^l(\lambda)$ into (\ref{eq:UB1}),  we obtain an upper bound on $q(\lambda)$
by an upper bound on $\mathbb{P}\left(Y^{\rm f}>1/\beta\right)$.  We presently give three such bounds, using the Markov and Chebychev inequalities and the Chernoff bound.  Although the details of the analysis below differ for each of the three bounds, the general techniques is the same: upper bound $\mathbb{P}\left(Y^{\rm f}>1/\beta\right)$ using the inequality, substitute into (\ref{eq:UB1}), then seek a simple expression that upper bounds the resulting expression.

The Markov inequality \cite{MitUpf2005} gives $\mathbb{P}(Y^{\rm f} > 1/\beta) \leq \beta \E[Y^{\rm f}]$.  Campbell's Theorem \cite{StoKen1996} states that if $\{ X_i \}$ are points drawn from a PPP of possibly varying intensity $\lambda(x)$ then
\begin{equation}
\mathbb{E} \left[\sum_{i \in \Pi} f(X_i) \right] = \int_{\mathbb{R}^2} f(x) \lambda( {\rm d}x).
\end{equation}
Applying this to find $\E[Y^{\rm f}]$ is straightforward after a change of variable to polar coordinates:
\begin{equation}
\label{eq:markov}
\E[Y^{\rm f}] = \mathbb{E} \left[ \frac{1}{r^{-\alpha}} \sum_{i \in \Pi \cap \bar{b}(0,s)} |X_i|^{-\alpha} \right] = r^{\alpha} \int_s^{\infty} t^{-\alpha} \lambda 2\pi t {\rm d} t = \frac{2\pi r^2 \beta^{\frac{2}{\alpha}-1}}{\alpha-2} \lambda \equiv \mu \lambda,
\end{equation}
where $s = \beta^{\frac{1}{\alpha}}r$.  Multiplying (\ref{eq:markov}) by $\beta$ and combining with \eqref{eq:UB1}, an upper bound on outage is
\begin{equation}
\label{eq:UB2}
q(\lambda) \leq q^{u,{\rm Markov}}(\lambda) = \left(1 - {\rm e}^{-\lambda \pi r^2 \beta^{\frac{2}{\alpha}}} \right) + {\rm e}^{-\lambda \pi r^2 \beta^{\frac{2}{\alpha}}} \frac{2\pi r^2 \beta^{\frac{2}{\alpha}}}{\alpha - 2} \lambda .
\end{equation}
Using the bounds $1-{\rm e}^{-A} \leq A$ and ${\rm e}^{-A} \leq 1$ for $A > 0$ and simplifying gives a ``relaxed Markov'' upper bound:
\begin{equation}
\label{eq:UB3}
q^{u,{\rm Markov}}(\lambda) \leq \pi r^2 \beta^{\frac{2}{\alpha}} \lambda  + \frac{2 \pi r^2 \beta^{\frac{2}{\alpha}}}{\alpha - 2} \lambda = \frac{\alpha}{\alpha-2} \pi r^2 \beta^{\frac{2}{\alpha}} \lambda .
\end{equation}
Setting \eqref{eq:UB3} equal to $\epsilon$ and solving for $\lambda$ gives a relaxed Markov lower bound on the TC:
\begin{equation}
\label{eq:UB4}
c^{l,{\rm Markov }}(\epsilon) = \frac{\alpha -2}{\alpha} \frac{\epsilon}{\pi r^2 \beta^{\frac{2}{\alpha}}}  + O(\epsilon^2) \mbox{ as } \epsilon \to 0,
\end{equation}
which is clearly smaller than the TC upper bound of \eqref{eqn:7} by a factor $(\alpha-2)/\alpha$.   The right hand side is obtained by observing that the first order Taylor series expansion of $\epsilon (1-\epsilon)$ around $\epsilon = 0$ equals $\epsilon + O(\epsilon^2)$.  Neglecting the $O(\epsilon^2)$ term gives an error $\epsilon - \epsilon (1-\epsilon) = \epsilon^2 = 0.01$ for $\epsilon = 0.1$.

Campbell's Theorem also gives the variance of the far-field aggregate interference:
\begin{equation}
\label{eqn:varint}
\mathrm{Var}(Y^{\rm f}) = \mathbb{E} \left[\frac{1}{r^{-2\alpha}} \sum_{i \in \Pi \cap \bar{b}(0,s)} \left(|X_i|^{-\alpha} \right)^2 \right] = \lambda r^{2\alpha} \int_s^{\infty} t^{-2\alpha} 2\pi t {\rm d} t =
\frac{\pi r^2 \beta^{\frac{2}{\alpha}-2}}{\alpha-1} \lambda \equiv \sigma^2 \lambda
\end{equation}
We use (\ref{eq:markov}) and (\ref{eqn:varint}) and Chebychev's inequality \cite{MitUpf2005} on the far-field aggregate interference (assuming $\mathbb{E}[Y^{\rm f}] < \frac{1}{\beta}$), as:
\begin{equation}
\label{eqn:Chebfar}
\mathbb{P} \left(Y^{\rm f} > \frac{1}{\beta} \right)
\leq \mathbb{P} \left(\left| Y^{\rm f} - \mathbb{E}[Y^{\rm f}] \right| > \frac{1}{\beta} - \mathbb{E}[Y^{\rm f}] \right)
\leq \frac{\sigma^2 \lambda}{\left( \frac{1}{\beta}- \mu \lambda \right)^2}
\end{equation}
Substituting (\ref{eqn:Chebfar}) into (\ref{eq:UB1}) and using the bounds $1-{\rm e}^{-A} \leq A$ and ${\rm e}^{-A} \leq 1$ for $A > 0$ and simplifying gives a ``relaxed Chebychev'' upper bound:
\begin{equation}
q^{u,{\rm Chebychev}}(\lambda) \leq \pi r^2 \beta^{\frac{2}{\alpha}} \lambda +  \frac{\frac{\pi r^2 \beta^{\frac{2}{\alpha}-2}}{\alpha-1} \lambda}
{\left( \frac{1}{\beta}- \frac{2\pi r^2 \beta^{\frac{2}{\alpha}-1}}{\alpha-2} \lambda \right)^2}.
\end{equation}
This expression is quadratic in $\lambda$; setting equal to $\epsilon$ and solving for $\lambda$ gives the relaxed Chebychev lower bound on the TC.

\begin{figure}
\centering
\includegraphics[width=3.5in]{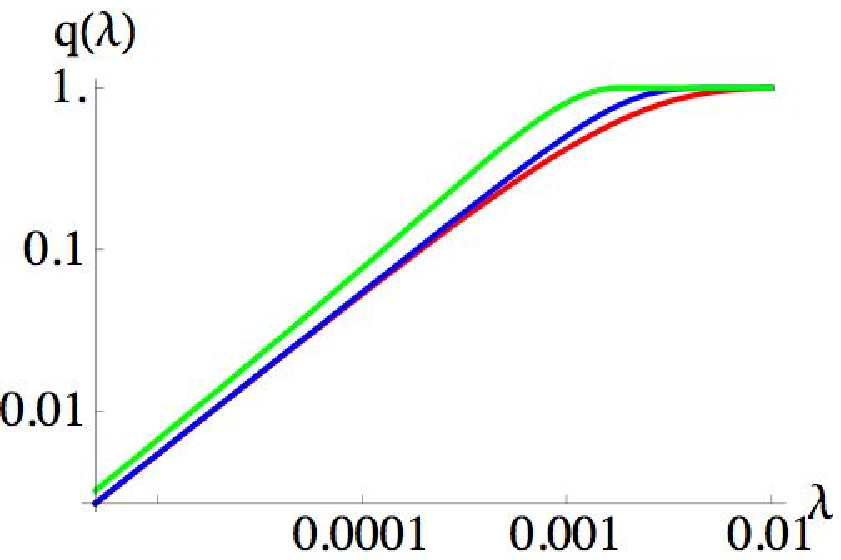}
\includegraphics[width=3.5in]{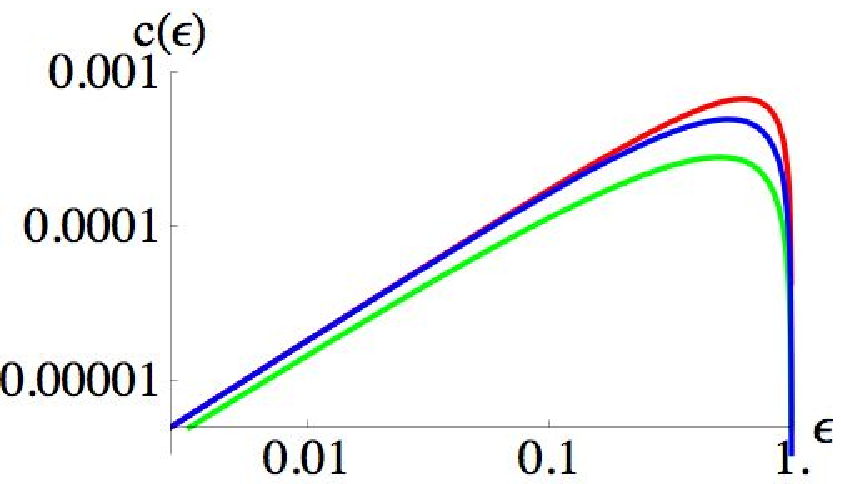}
\caption{{\bf Top:} OP $q(\lambda)$ versus the spatial intensity of attempted transmissions, $\lambda$, for the basic model with $\alpha = 4$, $\beta = 3$, and $r=10$ meters.  The three lines are lower bound, exact OP, and the (Chernoff) upper bound.  {\bf Bottom:} the TC $c(\epsilon)$ versus the outage requirement $\epsilon$ obtained by inverting the outage expression and bounds.}
\label{fig:b}
\end{figure}

The Chernoff bound \cite{MitUpf2005} may be used to obtain an upper bound on the OP:
\begin{equation}
\label{eqn:9}
\mathbb{P} \left(Y^{\rm f} > \frac{1}{\beta} \right) \leq \inf_{\theta \geq 0} \mathbb{E} \left[\ex^{\theta Y^{\rm f}} \right] \ex^{-\theta \frac{1}{\beta}} = \exp \left\{- \sup_{\theta \geq 0} \left(\theta \frac{1}{\beta} - 2 \pi \lambda \int_{\beta^{\frac{1}{\alpha}}r}^{\infty} \left(\ex^{\theta r^{\alpha} x^{-\alpha}} - 1 \right) x \dd x \right)  \right\}.
\end{equation}
This expression may be obtained by computing the moment generating function of $Y^{\rm f}$ restricted to $b(o,v)$ and then letting $v \to \infty$, as in \cite{SouSil1990,WebAnd2007a}.  The final upper bound on OP is then:
\begin{equation}
\label{eqn:10}
q^{u,{\rm Chernoff}}(\lambda) \equiv 1 - \left(1 - \exp \left\{- \sup_{\theta \geq 0} \left(\frac{\theta}{\beta} - 2 \pi \lambda \int_{\beta^{\frac{1}{\alpha}}r}^{\infty} \left(\ex^{\theta r^{\alpha} x^{-\alpha}} - 1 \right) x \dd x \right)  \right\}  \right) \ex^{-\lambda \pi r^2 \beta^{\frac{2}{\alpha}}}.
\end{equation}
Although the Chernoff OP upper bound is in some cases tighter than its Markov or Chebychev counterparts, it depends upon $\lambda$ in a complicated way which precludes a closed-form expression for the corresponding lower bound on the TC.  In this case, numerical inversion techniques must be applied.  Sample lower and upper bounds and exact expressions for both OP and TC are shown in Fig.~\ref{fig:b}.

\subsection{\label{ssec:subexp} Tightness of the lower bound: sub-exponential distributions}

Comparing the lower outage bound (\ref{eqn:6}) with the upper outage bound (\ref{eqn:10}), and glancing at Fig.~\ref{fig:b}, it is apparent that the (simple) lower outage bound is much tighter than the (complicated) upper bound.   One explanation for this comes from the fact that the random interference contribution of each node obeys a {\em subexponential distribution} \cite{GolKlu1997}.  Consider $n$ points distributed independently and uniformly over a disk of radius $d$ centered at the origin, denoted $\{X_1,\ldots,X_n\}$.  It is straightforward to establish the CCDF of the individual interference rvs, $V = |X|^{-\alpha}$, to be $\bar{F}_V(v) = \left(v^{\frac{1}{\alpha}} d \right)^{-2}$ for $v \geq d^{-\alpha}$.  A sufficient condition for a distribution to be subexponential is that $\lim \sup_{v \to \infty} v h_V(v) < \infty$  where $h_V(v) \equiv \frac{\dd}{\dd v}\left(-\log \bar{F}_V(v) \right)$ is the {\em hazard rate} function.  In our case, we find $v h_V(v) = \frac{2}{\alpha}$, ensuring $\bar{F}_V$ is subexponential.  A defining characteristic of subexponential distributions is the fact that sums of iid rvs $\{V_1,\ldots,V_n\}$ typically achieve large values $v$ by having one or more large summands (as opposed to a large number of moderate sized summands) \cite{GolKlu1997}:
\begin{equation}
\lim_{v \to \infty} \frac{\mathbb{P}(V_1+\cdots+V_n > v)}{\mathbb{P}(\max\{V_1,\ldots,V_n\} > v)} = 1, ~ n \geq 2.
\end{equation}
Because the interference contributions from each node are subexponential, it follows that the probability of an outage event $\{V_1+\cdots+V_n > v\}$ (for large $v$) approximately equals the probability of there being one or more dominant nodes with $V_i > v$.  Replacing $\sum_{i \in \Pi(\lambda)} |X_i|^{-\alpha}$ in (\ref{eqn:1}) with $\sum_{i=1}^n |X_i|^{-\alpha}$ gives $v = r^{-\alpha} \frac{1}{\beta}$.  Thus $v$ is large if {\em either} $\beta$ is small (receiver can decode small SIR) or $r$ is small (Tx and Rx are close together) .  For small $v$ (meaning {\em both} $\beta$ and $r$ are large), outage occurs more easily, and in particular, outage may occur due to the aggregate interference being large, even though there may not be any dominant nodes.  This argument holds for fixed $d$ and $n$, but gives intuition as to why the dominant interference lower bound is tight.

\subsection{\label{ssec:optimization} Optimization of SINR Threshold and Outage Constraint}

\begin{figure}
\centering
\includegraphics[width=3.5in]{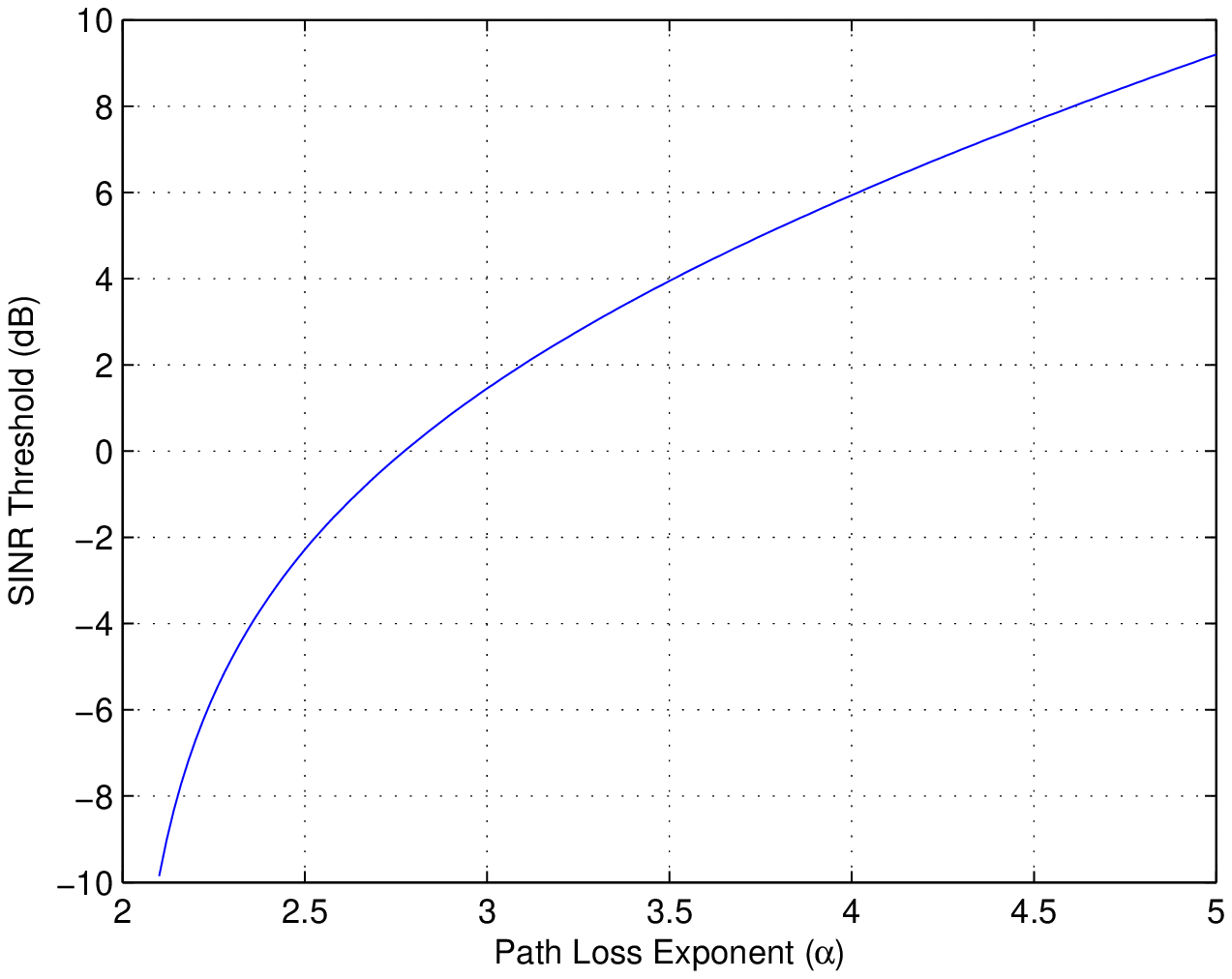}
\includegraphics[width=3.5in]{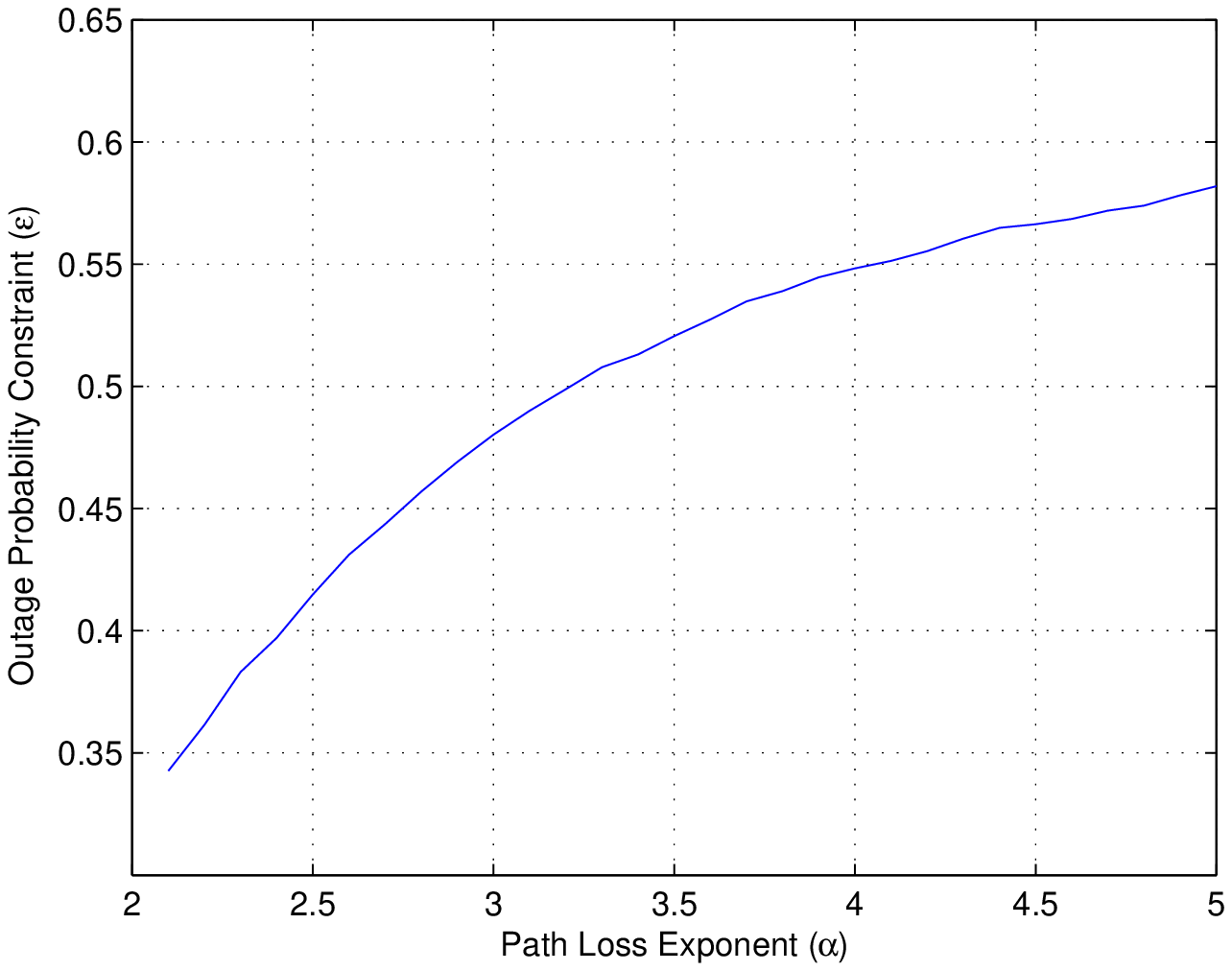}
\caption{{\bf Top:} Optimized SINR threshold $\beta$ versus path loss exponent $\alpha$.
{\bf Bottom:} Optimized outage probability constraint $\epsilon$ versus path loss exponent $\alpha$.}
\label{fig:opt}
\end{figure}

The SINR threshold $\beta$ and the outage constraint $\epsilon$, which are treated as constants in the TC framework, are generally under the control of the system designer and  should be chosen reasonably.  A meaningful objective is maximization of the area spectral efficiency $c(\epsilon) \log_2(1 + \beta)$, i.e., the product of successful density and spectral efficiency. Using (\ref{eq-transcap_implicit}), the joint maximization over $(\beta,\epsilon)$ can be written as:
\begin{align}
\max_{\beta, \epsilon} ~ c(\epsilon) \log_2(1 + \beta) &= \max_{\beta, \epsilon} ~
\frac{ \left(\bar{F}_{Z_{\alpha}}^{-1}(\epsilon)\right)^{-\frac{2}{\alpha}} (1 - \epsilon) }{\pi r^2 \beta^{\frac{2}{\alpha}}} \log_2(1 + \beta).
\end{align}
This clearly allows for separate maximizations of $\beta$ and $\epsilon$:
\begin{align}
\beta^\star = \textrm{arg} \max_\beta ~ \frac{\log_2(1 + \beta)}{\beta^{\frac{2}{\alpha}}}, ~~
\epsilon^\star = \textrm{arg} \max_\epsilon ~ \left(\bar{F}_{Z_{\alpha}}^{-1}(\epsilon) \right)^{-\frac{2}{\alpha}} (1 - \epsilon),
\end{align}
where the optimizers $\beta^\star$ and $\epsilon^\star$ depend only on the path-loss coefficient $\alpha$.  In \cite[Section IV]{JinAnd2008}, where a related but slightly different problem is studied, a closed-form solution for $\beta^\star$ was found:
\begin{equation}
\beta^\star = \mathrm{e}^{\frac{\alpha}{2} + \mathcal{W} \left( -
\frac{\alpha}{2} e^{-\frac{\alpha}{2}} \right)} - 1
\end{equation}
where $\mathcal{W}(z)$ is the principle branch of the Lambert $\mathcal{W}$ function.
$\bar{F}_{Z_{\alpha}}(\cdot)$ is not known in closed form, and thus $\epsilon^\star$ must be determined numerically.  In Fig.~\ref{fig:opt}, $\beta^\star$ and $\epsilon^\star$ are plotted versus $\alpha$, and both are seen to be increasing in $\alpha$. $\beta^\star$ is consistent with normal operating spectral efficiencies, while $\epsilon^\star$ shows that the optimal $\epsilon$ that maximizes the TC may be unacceptably large.  Although such a large outage provides a large area spectral efficiency, it also  translates directly to long transmission delays and energy inefficiency.  This analysis highlights a key drawback in unrestricted (spatial) throughput maximization: the max-throughput operating point may have an unacceptably high associated OP.  The TC framework captures this tradeoff by definition: it gives the maximum spatial throughput subject to a specified OP constraint.

\section{\label{sec:Design2} Transmission Capacity in Fading Channels}

We now evolve the discussion to consider channels that also have a random fluctuation about the path loss, commonly known as fading or shadowing.  The SIR in (\ref{eqn:1}) models the scenarios discussed in this section where random variable $S$ represents the desired signal fade and $I_i$ the fading coefficient from the $i$-th interferer.  We assume $S$ is drawn according to some distribution $F_S$ and each $I_i$ according to $F_I$ with $S, I_1, I_2, \ldots$ independent.  Independent fading is assumed for tractability; computing the OP and TC in correlated fading will be more difficult.

We first develop a framework for analyzing OP and TC with an arbitrary random channel, and then show exact results on OP and TC for Rayleigh and Nakagami fading.  It is initially surprising that exact results on OP and TC can be computed with certain types fading, but not without fading; recall in the previous session we had to be content with upper and lower bounds.  Although unmitigated fading reduces TC, it raises the possibility of opportunistic scheduling and transmit power control, which are discussed in \S \ref{ssec:scheduling} and \S \ref{ssec:FPC}.

\subsection{\label{ssec:fading} General Fading}

With general fading values as in (\ref{eqn:1}), the set of dominant interferers in (\ref{eqn:3}) becomes
\begin{equation}
\Pi^{\rm n}(\lambda) = \left\{ i : \frac{S r^{-\alpha}}{I_i |X_i|^{-\alpha}} < \beta \right\}.
\end{equation}
Computation of the probability of a dominant interferer ($\mathbb{P}(\Pi^{\rm n}(\lambda) \neq \emptyset)$) yields the following lower bound to OP \cite{WebAnd2007b}:
\begin{equation}
\label{eq-fading_general}
q^l(\lambda) = 1 - \mathbb{E} \left[ \exp \left\{- \lambda \pi r^2 \beta^{\frac{2}{\alpha}} \mathbb{E}[I^{\frac{2}{\alpha}}] S^{-\frac{2}{\alpha}}  \right\} \right],
\end{equation}
where the outer expectation is with respect to $S$.  This expression is similar to the LB in \eqref{eqn:6}, but the expectation in front of the exponential makes inverting this expression for $\lambda$ infeasible.  Applying Jensen's inequality to $q^l(\lambda)$ yields the following \emph{approximations}:
\begin{eqnarray}
\label{eq-op_approx_fading}
q(\lambda) &\approx& 1 - \exp \left\{- \lambda \pi r^2 \beta^{\frac{2}{\alpha}} \mathbb{E}[I^{\frac{2}{\alpha}}] \mathbb{E} [S^{-\frac{2}{\alpha}} ]  \right\} \\
\label{eq-tc_approx_fading}
c(\epsilon) &\approx& \frac{ -(1-\epsilon) \log (1-\epsilon)}{\pi r^2 \beta^{\frac{2}{\alpha}} \mathbb{E}[I^{\frac{2}{\alpha}}] \mathbb{E} [S^{-\frac{2}{\alpha}} ] }.
\end{eqnarray}
These quantities are approximations because Jensen's inequality yields inequality in the wrong direction. However, numerical results show that this approximation is reasonably accurate for small values of $\epsilon$ \cite{WebAnd2007b}. It is possible to extend the upper bounds from \S \ref{ssec:UB1} to fading \cite{WebAnd2007b}, but we focus exclusively on the above lower bound and approximation because they are more accurate.

If we assume that the signal and interference coefficients follow the same distribution $F_H$, which is reasonable in most communication environments, the expressions in (\ref{eq-fading_general})-(\ref{eq-tc_approx_fading}) particularize to:
\begin{eqnarray}
\label{eq-outage_lower_fading}
q^l(\lambda) &=& 1 - \mathbb{E}_{H} \left[ \exp \left\{- \lambda \pi r^2 \beta^{\frac{2}{\alpha}} \mathbb{E}[H^{\frac{2}{\alpha}}] H^{-\frac{2}{\alpha}}  \right\} \right] \\
q(\lambda) &\approx& 1 - \exp \left\{- \lambda \pi r^2 \beta^{\frac{2}{\alpha}} \mathbb{E}[H^{\frac{2}{\alpha}}] \mathbb{E} [H^{-\frac{2}{\alpha}} ]  \right\} \\
c(\epsilon) &\approx& \frac{ (1-\epsilon) \log (1-\epsilon)^{-1}}{\pi r^2 \beta^{\frac{2}{\alpha}} \mathbb{E}[H^{\frac{2}{\alpha}}] \mathbb{E} [H^{-\frac{2}{\alpha}} ] }.
\label{eq-tc_approx_fading2}
\end{eqnarray}
Comparing the TC approximation in (\ref{eq-tc_approx_fading2}) to the TC upper bound in (\ref{eqn:7}) we see that the effect of fading is captured by the term $ \left( \mathbb{E}[H^{\frac{2}{\alpha}}] \mathbb{E} [H^{-\frac{2}{\alpha}} ] \right)^{-1}$.  By Jensen's inequality, this quantity is less than one (with equality only if $H$ is deterministic) and thus fading has an overall negative effect relative to pure pathloss attenuation.  Furthermore, note that the TC approximation in (\ref{eq-tc_approx_fading2}) is equal to the exact TC in (\ref{eqn:ac}) for Rayleigh fading derived in the next section.  For the particular case of Rayleigh fading with $\alpha = 4$, the approximate ratio (\ref{eqn:7}) over (\ref{eqn:ac}) equals $\frac{\pi}{2} \approx 1.5708$, while the exact ratio ((\ref{eqn:12}) over (\ref{eqn:ac})) is $\frac{\pi}{2} \frac{Q^{-1}((1+\epsilon)/2)}{\log(1-\epsilon)^{-1}}$, which rapidly approaches $\frac{\pi}{2}$ as $\epsilon \to 0$.  Thus, adding Rayleigh fading to a network with $\alpha=4$ reduces the TC by $57\%$.

\subsection{\label{ssec:Rayleigh}Rayleigh Fading}

The case of Rayleigh fading, where each $H_{ij}$ is exponentially distributed (unit mean), is appealing not only for its practical importance but also because it is one of the few cases for which the OP and TC can be computed in closed form. The following argument was made precise by Baccelli {\em et al.} \cite{BacBla2006}, but can be traced to \cite{Lin1992,ZorPup1995}. Define the aggregate interference seen at the origin as $Z = \sum_{i \in \Pi(\lambda)}  H_{i0} |X_i|^{-\alpha}$, and denote the Laplace transform of $Z$ by $\mathcal{L}_Z(s) = \mathbb{E}\left[{\rm e}^{-s Z} \right]$. Then the success probability under Rayleigh fading is the Laplace transform of $Z$ evaluated at $s = \beta r^{\alpha}$:
\begin{equation}
\left. \mathbb{P}(\sir > \beta) = \mathbb{P}(H_{00} > \beta r^{\alpha} Z) = \int_0^{\infty} \ex^{-\beta r^{\alpha} z} f_Z(z) \dd z = \mathbb{E}\left[{\rm e}^{-s Z} \right] \right|_{s = \beta r^{\alpha}}.
\end{equation}
This transform can be computed explicitly, yielding an exact OP expression ((3.4) in \cite{BacBla2006}):
\begin{equation}
\label{eqn:ab}
q(\lambda) = 1 - \exp \left\{ - \lambda \pi r^2 \beta^{\frac{2}{\alpha}} \frac{2\pi}{\alpha} \csc \left(  \frac{2\pi}{\alpha} \right)  \right\},
\end{equation}
where $\csc$ denotes the cosecant.  The corresponding exact TC expression is
\begin{equation}
\label{eqn:ac}
c(\epsilon) = \frac{(1-\epsilon)\log(1-\epsilon)^{-1}}{\pi r^2 \beta^{\frac{2}{\alpha}} \frac{2\pi}{\alpha} \csc \left(  \frac{2\pi}{\alpha} \right) }.
\end{equation}

\subsection{\label{ssec:Nakagami}Nakagami Fading}

The Nakagami-$m$ distribution has power given by
\begin{equation}
f_S(x) = \left( \frac{m}{\E[S]}\right)^m \frac{x^{m-1}}{\Gamma(m)} \exp \left( -\frac{mx}{\E[S]} \right), ~ m \geq 0.5.
\end{equation}
and is quite general in that Rayleigh fading corresponds to $m = 1$ and path loss only corresponds to $m \to \infty$.  Because the distribution is also of exponential form, OP and TC can be computed exactly in a manner similar to Rayleigh fading, resulting in a transmission capacity of \cite{HunAnd2008a}
\begin{equation}
\label{eq:nak}
c(\epsilon)=\frac{K_{\alpha,m}(1-\epsilon)\log(1-\epsilon)^{-1}}{C_{\alpha,m}\beta^{\frac{2}{\alpha}}R^2}, ~~{\rm where}
\end{equation}
\begin{eqnarray}
K_{\alpha,m} &=& \left[1+\sum_{k=0}^{m-2}\frac{1}{(k+1)!} \prod_{l=0}^{k}(l-2/\alpha)\right]^{-1},\\
C_{\alpha,m} &=& \frac{2\pi}{\alpha}\sum^{m-1}_{k=0}{\binom{m}{k}} B\left(\frac{2}{\alpha}+k;m-\left(\frac{2}{\alpha}+k\right)\right),
\end{eqnarray}
and $B(a,b)=\frac{\Gamma(a)\Gamma(b)}{\Gamma(a+b)}$ is the Beta function.  Although this expression is clearly more complex than \eqref{eqn:ac}, it does describe nearly any fading environment.  Interestingly, if $m \to \infty$, i.e. for path loss only, \eqref{eq:nak} converges to the upper bound of \eqref{eqn:7}.

\subsection{\label{ssec:scheduling} Threshold scheduling}

Fading can potentially be exploited if only users experiencing good fading conditions transmit.  This can be done through a simple {\em threshold scheduling} rule where each transmitter elects to transmit only if the signal fading coefficient $H_{00}$ is larger than a threshold $t$, as in \cite{WebAnd2007b}.  Threshold scheduling is an example of opportunistic scheduling.  The spatial intensity of attempted transmissions for threshold $t$ is $\mu(t) \equiv \lambda \mathbb{P}(H_{00}> t) = \lambda \bar{F}_H(t)$, \ie, the original intensity $\lambda$ thinned by the probability of being above the threshold.  Because the threshold is on the received signal strength rather than the SIR, the decision depends only on local fading and does not affect the interference.  Therefore, the outage probability with threshold $t$ is:
\begin{equation}
\label{eq-thresh}
q(\nu,t) = \mathbb{P} \left(  \left. \frac{H_{00} r^{-\alpha}}{\sum_{i \in \Pi(\nu)} H_{i0} |X_i|^{-\alpha}} < \beta ~ \right| ~ H_{00} \geq t \right).
\end{equation}
where the $\{H_{ij}\}$ are drawn iid according to $F_H$.  The density of active transmissions is kept equal to $\nu$, independent of the value of $t$, by choosing $\lambda = \frac{\nu}{\mathbb{P}(H_{00}> t)}$. Thus, the only change brought about is that the signal distribution follows distribution $F_{H|H \geq t}$ instead of $F_H$. As a result, the OP in (\ref{eq-thresh}) is \textit{decreasing} in $t$ and thus TC \textit{increases} with $t$.\footnote{An outage is declared only if a transmitter actually attempts transmission and fails; not meeting the threshold is not considered an outage because it is essentially the same as not electing to transmit in pure Aloha.}  The transmission capacity approximation is given by:
\begin{eqnarray}
c(\epsilon) \approx \frac{ (1-\epsilon) \log (1-\epsilon)^{-1}}{\pi r^2 \beta^{\frac{2}{\alpha}} \mathbb{E}[H^{\frac{2}{\alpha}}] \mathbb{E} [H^{-\frac{2}{\alpha}} | H \geq t ] }.
\end{eqnarray}
Comparing this with (\ref{eq-tc_approx_fading2}), the (approximate) ratio of TC with threshold scheduling to that without it is $\frac{\mathbb{E} [H^{-\frac{2}{\alpha}}] }{ \mathbb{E} [H^{-\frac{2}{\alpha}} | H \geq t ]}$.  Because bad signal fades are eliminated, the gains from threshold scheduling can be very substantial: for example, in Rayleigh fading a very reasonable threshold of $t=1$ (i.e., $0$ dB) increases TC by a factor of $4.7$, $3.3$, and $2.25$ for $\alpha=2.5, 3,$ and $4$,
respectively.

\subsection{\label{ssec:FPC} Power control}

While threshold scheduling attempts to completely avoid bad fades, an alternative strategy is to transmit regardless of the fading conditions and adjust transmit power to compensate for fading.  In \cite{JinWeb2008} a \textit{fractional power control} policy in which each transmitter \emph{partially} compensates for the signal fading coefficient is proposed.  In particular, transmit power is chosen proportional to the fading coefficient raised to the exponent $-\gamma$ where $\gamma \in [0,1]$:
\begin{equation}
P_i^{\rm tx,fpc} = \frac{\rho}{\mathbb{E}[H_{ii}^{-\gamma}]} H_{ii}^{-\gamma} ~~~~ P_i^{\rm rx,fpc} = \frac{\rho}{\mathbb{E}[H_{ii}^{-\gamma}]} H_{ii}^{1-\gamma} r^{-\alpha}.
\end{equation}
Note that $\gamma=0$ corresponds to constant power while $\gamma=1$ corresponds to full channel inversion.  The resulting SIR is  $\sir = H_{00}^{1-\gamma} r^{-\alpha}/ \sum_{i \in \Pi(\lambda)} \left( H_{ii}^{-\gamma} H_{i0} \right) |X_i|^{-\alpha}$.

With channel inversion ($\gamma=1$) there is no signal fading ($S=1$) and each interference coefficient is distributed as $\frac{1}{H_{ii}}$, and thus based on (\ref{eq-fading_general}) we get the following OP lower bound:
\begin{equation}
q^{l,{\rm ci}}(\lambda) = 1 - \exp \left\{- \lambda \pi r^2 \beta^{\frac{2}{\alpha}} \mathbb{E}[H^{\frac{2}{\alpha}}] \mathbb{E}[H^{-\frac{2}{\alpha}}]  \right\}.
\label{eq-ci_lower}
\end{equation}
(There is no outer expectation because the signal fading coefficient is deterministic.)  By Jensen's inequality, this quantity is larger than the OP lower bound for constant power given (\ref{eq-outage_lower_fading}), and thus the lower bounds indicate that inversion degrades performance. For Rayleigh fading this ordering is precise: the OP lower bound with channel inversion in (\ref{eq-ci_lower}) is equal to the actual OP with constant power given in (\ref{eqn:ac}), and thus constant power is strictly superior to inversion in Rayleigh fading.

Although inversion worsens performance, partial compensation for fading can be beneficial.  If we consider general $\gamma$ and substitute the appropriate distributions for $S$ and $I$ in (\ref{eq-tc_approx_fading}), we get:
\begin{equation}
\label{eq-tc_approx_fpc}
c^{{\rm fpc}}(\epsilon,\gamma) \approx \frac{(1-\epsilon)\log(1-\epsilon)^{-1}}{\pi r^2 \beta^{\frac{2}{\alpha}}\mathbb{E}\left[H^{\frac{2}{\alpha}} \right] \mathbb{E}\left[H^{-\gamma\frac{2}{\alpha}} \right] \mathbb{E} \left[ H^{-(1-\gamma) \frac{2}{\alpha}} \right]}.
\end{equation}
This approximation is maximized by minimizing $\mathbb{E}\left[H^{-\gamma\frac{2}{\alpha}} \right] \mathbb{E} \left[ H^{-(1-\gamma) \frac{2}{\alpha}} \right]$  over $\gamma \in [0,1]$, and an application of H\"{o}lder's inequality yields $\gamma^* = 1/2$.  Although this only ensures that $\gamma=1/2$ is optimal for the TC approximation, results in \cite{JinWeb2008} confirm that $\gamma=1/2$ is also near-optimal for a wide range of reasonable parameter values.\footnote{An important exception to this is for large values of $\epsilon$, \ie, dense networks, in which case the optimum tends towards constant power ($\gamma=0$).}  Using $\gamma \gg \frac{1}{2}$ over-compensates for signal fading and leads to interference levels that are too high, while $\gamma \ll \frac{1}{2}$ leads to small interference levels but an under-compensation for signal fading.  The benefit of FPC is substantial for small values of $\epsilon$ and $\alpha$.  In Rayleigh fading, FPC increases TC by a factor of $2.1$ and $1.2$ for $\alpha=2.5$ and $\alpha=4$, respectively, for small $\epsilon$.

\section{\label{sec:MIMO} Multiple antennas}

The amplitude and phase of fading channels vary quite rapidly over space, with an approximate decorrelation distance of half a wavelength ($6$ cm at $2.5$ GHz).  This allows multiple suitably-spaced antennas to be deployed at both the transmitter and receiver to generate $N_t N_r$ Tx-Rx antenna pairs, where $N_t$ and $N_r$ are the number of transmit and receive antennas.  Considerable work has been done on multi-antenna systems (MIMO) in the past decade, well summarized by \cite{DigAlD2004,PauGor2003}, and such systems are now quite well understood and are central to all emerging high-data rate broadband wireless standards.  However, much less is known regarding the use of antennas in \textit{ad hoc} networks. In addition to providing diversity and spatial multiplexing benefits, multiple antennas also provide the ability to perform interference cancellation.  Recent analysis of MIMO systems using the TC framework allows us to evaluate these different antenna techniques, and provides a very optimistic picture of the benefit of MIMO in ad hoc networks.

\subsection{\label{ssec:diversity} Diversity}

Broadly defined, diversity techniques use TX and RX antennas to mitigate fading and increase the received SNR.  With maximum-ratio combining/transmission (MRC \& MRT), the transmitter and receiver apply weighting vectors at the antenna arrays based only on the Tx-Rx channel matrix.  If the TX and RX weight vectors are denoted by ${\bf t}_0$ and ${\bf r}_0$, respectively, and ${\bf H}_i$ denotes the $N_r \times N_t$ channel matrix from the $i$-th transmitter, then the SIR equation (\ref{eqn:1}) becomes:
\begin{equation}
\sir = \frac{|{\bf r}_0^{\dagger} {\bf H}_0 {\bf t}_0|^2 r^{-\alpha}}{ \sum_{i \in \Pi(\lambda)} |{\bf r}_0^{\dagger} {\bf H}_i {\bf t}_i|^2 |X_i|^{-\alpha}}.
\end{equation}
Choosing the TX and RX weights as the right/left singular vectors of the largest singular value of ${\bf H}_0$ results in the signal coefficient being equal to the square of this singular value, and thus boosts signal power by a factor between $\max\{ N_t,N_r \}$ and $N_t N_r$. With an appropriate application of (\ref{eq-tc_approx_fading}), this implies that the TC scales as \cite{HunAnd2008a}:
\begin{eqnarray}
O(\max\{N_t,N_r\}^{\frac{2}{\alpha}}) \leq c(\epsilon) \leq O((N_t N_r)^{\frac{2}{\alpha}}) \mbox{ as } N_t,N_r \to \infty.
\end{eqnarray}
The upper bound is tight for channels with high spatial correlation, while the lower bound is tight for i.i.d. Rayleigh fading.  Note that $N_t=1, N_r > 1$ and $N_t > 1, N_r=1$ correspond to maximum-ratio combining (MRC) and maximum-ratio transmission (MRT), respectively.

Orthogonal space-time block coding (OSTBC) is another diversity technique.  OSTBC, which intuitively corresponds to repeating each information symbol from different antennas at different times, does not change the transmitted symbol rate but significantly increases received signal power.\footnote{For some combinations of $N_t$ and $N_r$ OSTBCs either lose orthogonality, or reduce the data rate slightly. The results here make the optimistic assumption of rate 1 orthogonal STBCs for general $N_t, N_r$.} However, interference power is also boosted and as a result OSTBCs increase the TC scaling only as $c(\epsilon) = O(N_r^{\frac{2}{\alpha}})$ \cite{HunAnd2008a}. OSTBCs have very little affect on TC -- the scaling gain is due to MRC at the receiver, independent of the code.

\subsection{\label{ssec:cancel} Spatial Interference Cancellation}

If the receiver also has knowledge of the interferer channels, the $N_r$-dimensional RX weight vector can be used to cancel interference.  In the single-transmit, multi-receive antenna setting with spatially uncorrelated Rayleigh fading, choosing the RX weight vector orthogonal to the vector channels of the strongest $N_r - 1$ interferers (i.e., ${\bf r}_0 \perp {\bf H}_1, \ldots, {\bf H}_{N_r - 1}$) results in $O( {N_r}^{1 - \frac{2}{\alpha}} )$ TC scaling \cite{HuaAndSub}.  An even larger TC increase is obtained if the RX vector is designed to cancel interference and reap diversity.  In particular, using about half the RX degrees of freedom for cancellation and the remainder for diversity (i.e.,  choosing ${\bf r}_0$ as the projection of vector ${\bf H}_0$ on the nullspace of ${\bf H}_1, \ldots, {\bf H}_{N_r/2}$ ) leads to $O(N_r)$ TC scaling \cite{JinAnd2009}.\footnote{Both of these scaling results are obtained using the OP upper bounding techniques described in \S \ref{ssec:UB1}.}  In fact, the SIR is maximized, and thus the benefits of interference cancellation and diversity are optimally balanced, if the RX vector is chosen according to the MMSE-criterion: ${\bf r}_0 = \left( \sum_{i \in \Pi(\lambda)} |X_i|^{-\alpha} {\bf H}_i {\bf H}_i^{\dagger} \right)^{-1/2} {\bf H}_0$.  The MMSE filter is generally quite difficult to deal with analytically, although large-system results are derived using random matrix theory in \cite{GovBil2007}.

\begin{figure}
\centering
\includegraphics[width=3.5in]{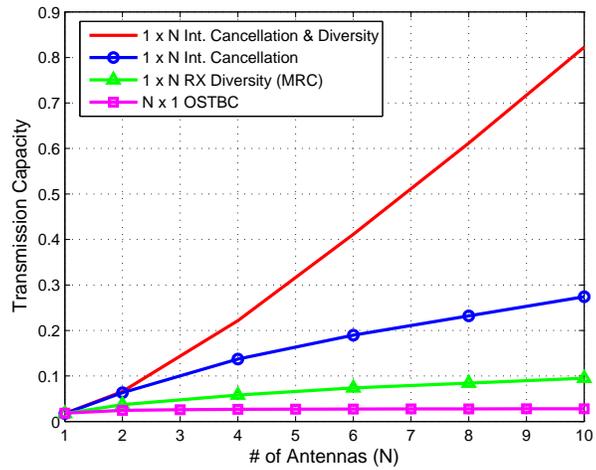}
\caption{The transmission capacity of various spatial diversity techniques vs. the number of antennas per node.  Interference
cancellation \& diversity refers to cancelling the nearest $N/2$ interferers and using the remaining degrees of freedom for diversity.
Here, $\epsilon = .1$, $\alpha = 4$, $\beta = 1$.}
\label{fig:diversity}
\end{figure}

In Fig. \ref{fig:diversity} the TC of diversity (beamforming and OSTBC) and intererence cancellation are plotted versus the number of antennas ($N$) for $\alpha=4$ and $\beta=1$.  All of the techniques except OSTBC provide significant gains, but the combination of interference cancellation and diversity clearly provides the largest TC, as predicted by the TC scaling results.

\subsection{\label{ssec:mux} Spatial Multiplexing}

\begin{figure}
\centering
\includegraphics[width=3.5in]{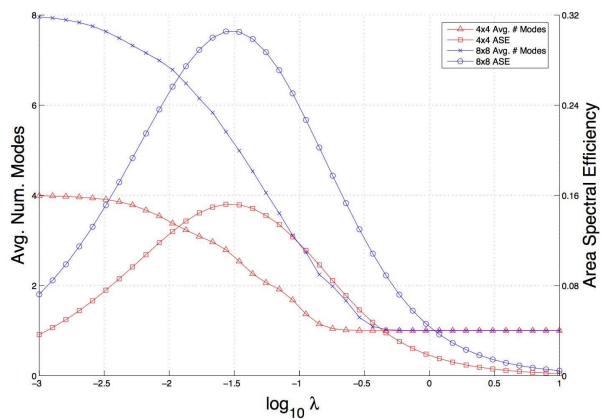}
\caption{Optimal number of MIMO modes $L$ and Area Spectral Efficiency (ASE) vs. transmitter intensity per m$^2$.  The $L$ curves are monotonically decreasing, ASE curves are bell-shaped and have a unique maximum.  Here, $\epsilon = .1$, $r=1$m, $\alpha = 4$.}
\label{fig:MIMO}
\end{figure}

The most aggressive use of the antennas is to use them to form up to $L \leq \min \{N_t, N_r \}$ parallel spatial channels. If the transmitter has knowledge of the channel matrix ${\bf H}_0$, this corresponds to beamforming along the eigenmodes of the channel.  The achieved SINR for each spatial channel depends on the eigenvalues of the channel matrix as well as the interference power, so some channels are much better than others. When subject to an SINR target and an outage constraint, it is preferable to transmit only a small number of streams ($L \ll N$) unless the network is very sparse. This is illustrated in Fig. \ref{fig:MIMO} where the optimized number of spatial streams
(as determined in \cite{HunAnd2008b}) is plotted versus the interferer density and this quantity is seen to decrease from $N$ to $1$ with
the density.  Ideally, the number of spatial channels can be adapted dynamically based on the channel and interference strengths to maximize the quantity $L c(\epsilon,L)$, which is the area spectral efficiency (ASE) shown in Fig. \ref{fig:MIMO}, and has a unique maximum \cite{HunAnd2008b}.  Here $c(\epsilon, L)$ is the TC with target OP $\epsilon$ when $L$ antennas are employed.  If each TX wishes to communicate with multiple receviers, \textit{multi-user} MIMO techniques can be used to send separate data streams to each receiver.  In the situation where each transmitter and receiver has $N$ antennas, the TC has been shown to increase \textit{super-linearly} with $N$ when dirty paper coding, the optimal multi-user MIMO technique, is used \cite{KouAnd2009a}.

If the transmitter does not know channel matrix ${\bf H}_0$, spatial multiplexing is generally performed by transmitting independent data streams from each transmit antenna.  The OP and TC for low-complexity (and sub-optimal) MRC and zero-forcing receivers are known \cite{LouMcK2009}, but many important questions remain unanswered on this topic, e.g., performance with optimal MIMO receivers.

\section{\label{sec:future} Current Limitations and Future Directions}

Although the results presented in this paper have illustrated the value of the transmission capacity framework, they have also failed to capture two important aspects of \adhoc networks.  The first is that they are for a snapshot, or single-hop, of the network.  This may be acceptable for unlicensed spectrum analysis or other decentralized networks, but {\em ad hoc} networks must route traffic from source to destination, often over multiple hops through intermediate nodes.  A network with higher single-hop TC should be able to achieve higher end-to-end capacity than a network with smaller TC because more simultaneous transmissions are possible. However, important issues such as desired hop length, number of hops, multi-hop routes, and end-to-end delay are not presently addressed.  In addition, noise should not be neglected since a principle function of multihop is to increase the SNR for each hop. Some work that attempts to use the results of this paper (or similar results) to address multihop includes \cite{BacBla2006}, where a metric called \emph{expected forward progress} is introduced and used to find the optimum split between transmitters and receivers (potential relays) in terms of the Aloha contention probability.  Recently, \cite{StaRos2009} has developed a multihop model and found an end-to-end delay-optimizing strategy in a Poisson field of interference (without noise), while \cite{AndWeb09} finds the end-to-end transmission capacity in closed-form (i.e., transport capacity) with noise under a few restrictive assumptions like equi-distant relays and independent retransmissions.  Clearly, this is a line of work that should be pursued and improved upon in the coming years.

The second lacking aspect of the current results is that they rely on a homogeneous Poisson distribution of nodes for tractability, which accurately models only uncoordinated transmissions (\eg, Aloha).   A well known alternative is to schedule simultaneous transmissions with the objective of controlling interference levels.  Local scheduling mechanisms generally space out simultaneous transmissions, thereby significantly changing the interference distribution, while idealized centralized scheduling can eliminate outages altogether and determine the optimal set of transmitters in each slot (\eg, max-weight scheduling within the backpressure paradigm \cite{GeoNee2006}).  Preliminary work in this direction includes computing the outage probability and transmission capacity under non-Poisson point processes \cite{HaeGanNOW,GanHae09,YanDeV2007}.  Although scheduling mechanisms provide obvious gains, these come at the cost of overhead (\eg, control messages). Thus, a general open question is understanding the tradeoff between the benefits and overhead costs of different scheduling/routing mechanisms (Aloha is a particular point on this tradeoff curve), and determining the appropriate techniques for different network settings.  Furthermore, a fundamental property that applies even to scheduled systems is that transmissions occupy space whenever interference is treated as noise; the transmission capacity provides a clean characterization of this space, and thus many of the insights apply, in principle, to scheduled systems as well.

As is true of any complicated research topic, discussion of a particular model or framework exposes tension between analytical tractability and accuracy/generality. The transmission capacity framework clearly leans towards simplicity and tractability, but nonetheless provides valuable design insight and a launching point for more refined, less tractable network analyses.

\section{Acknowledgements}
The authors appreciate the technical contributions of G. de Veciana, A. Hasan, A. Hunter, X. Yang, and K. Huang, which led to some of the results summarized here.  They also are grateful for feedback from M. Haenggi.

\bibliographystyle{IEEEtran}
\bibliography{TCTutorial}

\newpage

\begin{table}
\centering \caption{Notation used in paper.}
\begin{tabular}{|r|l|} \hline
$a \equiv b$ & $a$ is defined to equal $b$ \\
$\lambda$ & spatial intensity of attempted transmissions (per $m^2$) \\
$\Pi = \{X_i\}$ & Poisson point process (PPP) of intensity $\lambda$ of transmitter locations \\
$\alpha$ & pathloss exponent ($\alpha > 2$) \\
$\beta$ & SIR/SINR requirement for successful reception \\
$r$ & distance separating each Tx-Rx pair \\
$q(\lambda)$ & outage probability (OP) \\
$\epsilon$ & constraint on OP \\
$c(\epsilon)$ & transmission capacity (TC) \\
$\rho$ & transmission power \\
$H_{ij}$ & fading coefficient from transmitter $i$ to receiver $j$ \\
$M$ & number of frequency channels, or spreading factor \\
$N_r, N_t, N$ & number of receive, transmit, or total antennas \\
\hline
\end{tabular}
\label{tab:1}
\end{table}

\end{document}